\documentclass[12pt]{article}

\renewcommand{\d}{{\rm d}}
\newcommand{\gsth}{g_{\sigma^3}}
\newcommand{\sgn}{{\rm sgn}}
\newcommand{\Dsth}{\Delta_{\sigma^3}}
\newcommand{\sth}{\sigma^3}
\newcommand{\Msth}{M_{\sigma^3}}
\newcommand{\Vsth}{V_{\sigma^3}}
\newcommand{\bx}{\mbox{\boldmath$x$}}
\newcommand{\gsigd}{g_{\sigma^d}}
\newcommand{\Msigd}{M_{\sigma^d}}
\newcommand{\sigd}{\sigma^d}
\newcommand{\Dsigd}{\Delta_{\sigma^d}}
\newcommand{\Vsigd}{V_{\sigma^d}}
\newcommand{\ltwosigd}{l^2_{\sigma^d}}
\newcommand{\ltwosth}{l^2_{\sigma^3}}
\newcommand{\stw}{\sigma^2}
\newcommand{\son}{\sigma^1}
\newcommand{\sfo}{\sigma^4}
\newcommand{\ltwoson}{l^2_{\sigma^1}}
\newcommand{\Dsthk}{\Delta_{\sigma^3_k}}
\newcommand{\ltwosonsfok}{l^2_{\sigma^1\sigma^4_k}}
\newcommand{\ltwosonsfokplon}{l^2_{\sigma^1\sigma^4_{k+1}}}
\newcommand{\dthstwnul}{\delta^3_{\sigma^2}(0)}
\newcommand{\dsonnul}{\delta_{\sigma^1}(0)}
\newcommand{\Dstw}{\Delta_{\sigma^2}}
\newcommand{\gstw}{g_{\sigma^2}}

\newcommand{\dfuns}{$\delta$-functions }
\newcommand{\dsigd}{\delta_{\sigma^d}}
\newcommand{\Vstw}{V_{\sigma^2}}
\newcommand{\Vson}{V_{\sigma^1}}

\begin{document}

\title{Gravity action on discontinuous metrics}
\author{V.M. Khatsymovsky \\
 {\em Budker Institute of Nuclear Physics} \\ {\em
 Novosibirsk,
 630090,
 Russia}
\\ {\em E-mail address: khatsym@inp.nsk.su}}
\date{}
\maketitle
\begin{abstract}
We consider minisuperspace gravity system described by piecewise flat metric discontinuous on three-dimensional faces (tetrahedra). There are infinite terms in the Einstein action. However, starting from proper regularization, these terms in the exponential of path integral result in  pre-exponent factor with $\delta$-functions requiring vanishing metric discontinuities. Thereby path integral measure in Regge calculus is related to path integral measure in Regge calculus where length of an edge is not constrained to be the same for all the 4-tetrahedra containing this edge, i.e. in Regge calculus with independent 4-tetrahedra. The result obtained is in accordance with our previous one obtained from symmetry considerations.
\end{abstract}

PACS numbers: 04.60.-m Quantum gravity

\newpage

Regge calculus (RC) \cite{Regge} is general relativity for piecewise flat Riemannian metrics, i. e. those ones which are flat everywhere with exception f the set of zero measure. These manifolds can be viewed as collection of flat 4-dimensional tetrahedra or 4-simplices fitted to each other on their common 3-faces (usual tetrahedra). The curvature distribution residues on the 2-faces, i. e. triangles or 2-simplices $\sigma^2$. Metric can be chosen affine in each of the 4-simplex $\sigma^4$. When passing through a 3-face $\sigma^3$ metric generally changes, but only its normal to $\sigma^3$ component. The lengths of the 3-face $\sigma^3$ should be unambiguous when we pass from one 4-simplex to another one, so tangential metric should not suffer from discontinuity. Recently the so-called area RC \cite{BarRocWil,RegWil} is of interest where continuity of metric is violated \cite{WaiWil}. This is because number of 2-faces is larger than the number of edges $\sigma^1$. Therefore edge lengths should generally be ambiguous. The interest to area RC is provided by an idea that areas of the triangles are more natural than the edge lengths \cite{Rov}. In \cite{Kha} we have considered general case when there is no any relations connecting lengths of the same edge when passing from one 4-tetrahedron to another one. (Whereas area RC implies that areas of the same triangle be the same in the different 4-tetrahedra containing this triangle.) Suggesting to interpret this as discontinuity of (tangential component of) metric we have found the path integral measure in RC in terms of that in RC with independent 4-tetrahedra and thus ambiguous edge lengths. This has been done by projecting in configuration superspace the measure in the RC with independent 4-tetrahedra onto hypersurface corresponding to ordinary RC. The arising $\delta$-function factor needed to provide vanishing the metric discontinuities has being fixed by symmetry considerations. Now we would like to accurately determine the factor following from $\exp (iS_g)$ upon direct substitution of the (properly regularized) discontinuous metric into Einstein action $S_g$.

We consider Minkowsky metric $g_{\lambda\mu}$ with signature $(-,+,+,+)$, $\lambda$, $\mu$, \dots = 0,1,2,3. It is assumed constant in each 4-tetrahedra $\sigma^4$. Let $x^n$ be coordinate normal to the considered 3-face $\sigma^3$. The metric is chosen as $\d s^2 = g_{nn}(\d x^n)^2 + g_{\alpha\beta}\d x^{\alpha} \d x^{\beta}$. Here indices $\alpha$, $\beta$, \dots take on 3 values, $n$ is 4th index. Denote matrix $\| g_{\alpha\beta} \|$ for given $\sigma^3$ as $\gsth$. The 3-face $\sigma^3$ can be timelike or spacelike one, $\sgn \det \gsth$ = $- \sgn g_{nn}$. The $\gsth$ undergoes jump $\Dsth \gsth$ (or simply $\Delta \gsth$) across $\sth$. Let us approach to this via dependence on $x^n$ close to stepwise one which is linear in the interval $\Delta x^n \to 0$. Singular term in $S_g$ is
\begin{equation}
S_g = -{1\over 64 \pi G} \int (g^{\alpha\gamma}g^{\beta\delta} - g^{\alpha\beta}g^{\gamma\delta})g_{\alpha\beta , n}g_{\gamma\delta , n}g^{nn}\sqrt{-\det g}\d^4x.
\end{equation}

\noindent It will be convenient in what follows write out appearing here matrix of bilinear in $g_{\alpha\beta ,n}$ form as
\begin{equation}\label{M}
M^{(\alpha\beta)(\gamma\delta)} = {1\over 2}(g^{\alpha\gamma}g^{\beta\delta} + g^{\alpha\delta}g^{\beta\gamma}) - \lambda g^{\alpha\beta}g^{\gamma\delta}
\end{equation}

\noindent slightly generalizing it to $\lambda \neq 1$, $\lambda \to 1$. For certain $\sth$ denote it as $\Msth$. It is also important to define {\it dimensionless} regularisation parameter $\varepsilon$ as
\begin{equation}\label{varepsilon}
{1\over \varepsilon} = {6|\Vsth |\sgn \det \gsth \over 64\pi \Delta x^n \sqrt{|g_{nn}|}}.
\end{equation}

\noindent Here $6|\Vsth| = \Delta^3\bx \sqrt{|\det \gsth |}$ is 6 $\times$ module of volume of $\sth$. So we get factors in the path integral
\begin{equation}\label{exp3d}
\exp \left [ {i\over \varepsilon} (\Delta \gsth \Msth \Delta \gsth ) \right ].
\end{equation}

\noindent Exponential reminds simplicial DeWitt supermetric on 3-geometries (see, e. g., \cite{HamWil}). Also it will be of interest to know result of reduction in (\ref{exp3d}) of 3D metric to 1D or 2D subspace, i. e.
\begin{eqnarray}
g_{\alpha\beta} = \left( \begin{array}{cc} \begin{array}{c} d \times d \\ \mbox{block} \end{array} & 0 \\ 0 & \begin{array}{c} (3-d) \\ \times (3-d) \\ \mbox{block} \end{array} \end{array} \right), ~~~ \Delta g_{\alpha\beta} = \left( \begin{array}{cc} \begin{array}{c} d \times d \\ \mbox{block} \end{array} & 0 \\ 0 & 0 \end{array} \right)
\end{eqnarray}

\noindent Thus generalize notation (\ref{exp3d}) to
\begin{equation}
\exp \left [ {i\over \varepsilon} (\Delta \gsigd \Msigd \Delta \gsigd ) \right ]
\end{equation}

\noindent where at $d$ = 1,2 index $\sigd$ means reduction of $M$, $\Delta g$ to these subspaces.

We issue from
\begin{equation}
\exp (ix^2/\varepsilon ) = \sqrt{\pi\varepsilon} \exp (i\pi/4) \delta (x)
\end{equation}

\noindent at $\varepsilon \to 0$ as distribution on standard set of probe functions. Also
\begin{equation}
\det \Msigd = 2^{-d(d-1)/2}(1 - \lambda d) (\det \gsigd )^{-d-1}
\end{equation}

\noindent and eventually
\begin{eqnarray}\label{expd}
& & \exp \left [ {i\over \varepsilon} (\Delta \gsigd \Msigd \Delta \gsigd ) \right ] = \nonumber \\& & (1 - \lambda d)^{-1/2} [\sqrt{\pi\varepsilon} \exp (i\pi/4)]^{d(d+1)/2} 2^{d(d-1)/4} (\det \gsigd )^{(d+1)/2} \delta^{d(d+1)/2} (\Delta \gsigd ).
\end{eqnarray}

\noindent Here we can pass from $\gsigd$ to set of linklengths squared $l^2_{\sigd}$. In particular, $l^2_{\son}$ is ${\rm (length)}^2$ of the edge $\son$.
\begin{equation}
(\det \gsigd )^{(d+1)/2} \delta^{d(d+1)/2} (\Dsigd \gsigd ) = (d!\Vsigd )^{d+1} \delta^{d(d+1)/2} (\Dsigd \ltwosigd ),
\end{equation}

\noindent $\Vsigd$ is the volume of $\sigd$. We should take product of these factors at $d$ = 3 over $\sigd$ = $\sth$,
\begin{equation}\label{prodsth}
\prod_{\sth} \Vsth^4 \delta^6 (\Dsth \ltwosth).
\end{equation}

\noindent This contains extra $\delta$-functions and thus $\delta$-functions of zero. These are connected with cycles around triangles $\stw$. Given a link $\son$ and a
triangle $\stw \supset \son$, the lengths squared of $\son$ in the 4-tetrahedra $\sfo_1 , \dots , \sfo_n \supset \stw$ enter the product (\ref{prodsth}) through the product of $\delta$-functions of discontinuities of $\ltwoson$ on 3-faces $\sth_k = \sfo_k \cap \sfo_{k+1}$ between these 4-tetrahedra, $\Dsthk \ltwoson = \ltwosonsfok - \ltwosonsfokplon$, $k$ = 1,2, \dots, $n$ ($\sfo_{n+1} \equiv \sfo_1$. This product is not well-defined for it contains $\delta$-function of zero,
\begin{equation}
\prod^n_{k=1} \delta (\ltwosonsfok -\ltwosonsfokplon ) = \delta (0) \prod^n_{k=1}\! ^{\prime} \delta (\ltwosonsfok -\ltwosonsfokplon ).
\end{equation}

\noindent Here $\prod^{\prime}$ is well-defined product of $\delta$-functions obtained by omitting any one of the factors $\delta (\Dsthk \ltwoson)$. There are two else links forming the given triangle $\stw$ so we get $\dthstwnul \equiv \delta^3 (\Dstw \gstw = 0)$ in (\ref{prodsth}). Here subscript $\stw$ indicates that regularized $\delta^3 (\Dstw \gstw )$ at $\Dstw \gstw = 0$ depends on $\stw$ as it is seen from (\ref{expd}) at $d$ = 2, $\Dstw \gstw = 0$ read from right to left. If, however, we take product of $\dthstwnul$ over all $\stw$ sharing the link $\son$, this set of \dfuns of zero is larger than that contained in (\ref{prodsth}). Indeed, let $N$ be number of the triangles $\stw \supset \son$. Let $\sth_k$, $k = 1,2, \dots , N-1$ be chain of $N-1$ tetrahedrons successively passing through these $\stw$. That is, if numbered accordingly, $\stw_k = \sth_{k-1} \cap \sth_k$, $k = 2,3, \dots , N-1$, and $\stw_1 \subset \sth_1$ and $\stw_N \subset \sth_{N-1}$ are ending triangles in this chain. Then the product of $\delta (\Dsth \ltwoson )$ over $\sth \supset \son$, $\sth \not\in \{\sth_k | k = 1,2, \dots, N-1\}$ can be well-defined in (\ref{prodsth}). Only remaining $N-1$ \dfuns $\delta (\Dsthk \ltwoson)$, $k = 1,2, \dots, N-1$ acquire zero arguments, whereas product of $\dthstwnul$ over $\stw \supset \son$ contains just $N$ such functions. The result of throwing away extra factor $\dsonnul$ can be written as
\begin{equation}
(\dsonnul )^{-1}\prod_{\stw \supset \son} \dthstwnul.
\end{equation}

\noindent Considering this for all the links $\son$, we can write (\ref{prodsth}) as
\begin{equation}\label{prodprime}
\prod_{\sth} \Vsth^4 \left ( \prod_{\son,\sth \supset \son}\!\!\!\!\!\!^{\prime} \delta (\Dsth \ltwoson) \right )\prod_{\stw} \dthstwnul \left( \prod_{\son} \dsonnul \right )^{-1}.
\end{equation}

\noindent The primed product here means that redundant \dfuns are omitted (i. e. it is well-defined).

The $\dsigd^{d(d+1)/2} (0)$ follows from (\ref{expd}) read from right to left,
\begin{equation}\label{delta0}
\dsigd^{d(d+1)/2} (0) \propto \Vsigd^{d+1}.
\end{equation}

\noindent Reduction of matrix $M$ to 1-dimensional subspace is zero, and this is just the reason why do we have introduced $\lambda \neq 1$ in (\ref{M}) for the intermediate regularization. The factor $(1 - \lambda )^{-1/2}$ in (\ref{expd}) can be included into overall normalization factor for measure. Then we can pass to the limit $\lambda \to 1$. Substituting into (\ref{prodprime}) we get
\begin{equation}
\prod_{\sth}\Vsth^4 \prod_{\stw}\Vstw^{-3} \prod_{\son}\Vson^2 \prod_{\son,\sth \supset \son}\!\!\!\!\!\!^{\prime} \delta (\Dsth \ltwoson).
\end{equation}

The symmetry of this result (w.r.t. the different simplices) is due to the above choice of dimensionless $\varepsilon$ (\ref{varepsilon}) (which absorbs $\Vsth$) as regularization parameter universal for different 3-faces $\sth$. Were it not so, above $\dsigd^{d(d+1)/2} (0)$, $d = 1,2$ would depend not only on $\sigd$ (\ref{delta0}), but also on the choice of $\sth \supset \sigd$, and the considered symmetry w.r.t. the different simplices would not be achievable.

The question may arise why can not we do the same for any field different from gravity. Peculiar feature of gravity is that physical curvature distribution has 2-dimensional support while "virtual" curvature responsible for independence of neighboring 4-tetrahedra develops in the interior of 3-faces, that is, live at another points. It would be quite unnatural to suggest that effect of these points of 3-faces depends on another points, those of boundary 2-faces (i. e. that $\varepsilon$ depends on defect angles). Contrary to that, physical non-gravity field such as electromagnetic one, is distributed everywhere over 4-volume, including points of 3-faces. Therefore corresponding $\varepsilon$ well may depend on characteristics of this field. The problem of specifying $\varepsilon$ and thus the measure would require additional assumptions in this case.

The result obtained defines factor which connects path integral measure of genuine RC $\d \mu_{\rm Regge}$ with the measure corresponding to the action on independent 4-tetrahedra but without terms singular at $\varepsilon \to 0$, i. e. to the action which allows discontinuous metrics. If the action on independent 4-tetrahedra is the sum of independent parts referring to different 4-tetrahedra, the latter measure is evidently the product of those ones for separate 4-tetrahedra $\d \mu (\sfo)$,
\begin{equation}
\d \mu_{\rm Regge} = \prod_{\sfo} \d \mu (\sfo) \prod_{\sth}\Vsth^4 \prod_{\stw}\Vstw^{-3} \prod_{\son}\Vson^2 \prod_{\son,\sth \supset \son}\!\!\!\!\!\!^{\prime} \delta (\Dsth \ltwoson).
\end{equation}

The present work was supported in part by the Russian Foundation for Basic Research
through Grant No. 08-02-00960-a.


\end{document}